\documentclass[preprint,aps,pre,longbibliography]{revtex4-1}
\usepackage[utf8]{inputenc}
\usepackage{epsfig}
\usepackage{subfigure}
\usepackage{latexsym}
\usepackage{amsmath}
\usepackage{psfrag}
\usepackage{fullpage}
\usepackage{graphicx}
\usepackage{amssymb}
\usepackage{color}
\usepackage{placeins}
\usepackage{units}
\usepackage{makecell}
\usepackage{hyperref}

\DeclareMathOperator{\argmin}{argmin}

\newcommand{\avg}[1]{\left\langle #1\right\rangle}

\newcommand{\T}[1]{\textrm{#1}}

\begin{document}
 \title{Consensus ranking for multi-objective interventions in multiplex networks}

\author{M\'arton P\'osfai}
\email{posfai@ucdavis.edu}
\affiliation{Complexity Sciences Center and Department of Computer Science, University of California, Davis, CA 95616, USA}

\author{Niklas Braun}
\affiliation{Department of Mechanical and Aerospace Engineering, University of California, Davis, CA 95616, USA}

\author{Brianne A. Beisner}
\affiliation{Department of Population Health and Reproduction, University of California, Davis, CA 95616, USA}
\affiliation{Neuroscience and Behavior Unit, California National Primate Research Center, University of California, Davis, CA 95616, USA}

\author{Brenda McCowan}
\affiliation{Department of Population Health and Reproduction, University of California, Davis, CA 95616, USA}
\affiliation{Neuroscience and Behavior Unit, California National Primate Research Center, University of California, Davis, CA 95616, USA}

\author{Raissa M. D'Souza}
\affiliation{Complexity Sciences Center, Department of Computer Science and Department of Mechanical and Aerospace Engineering, University of California, Davis, CA 95616, USA}
\affiliation{{Santa Fe Institute, 1399 Hyde Park Road, Santa Fe, NM 87501,  USA}}

\date{\today}

\begin{abstract}
High-centrality nodes have disproportionate influence on the behavior of a network; therefore controlling such nodes can efficiently steer the system to a desired state.
Existing multiplex centrality measures typically rank nodes assuming the layers are qualitatively similar. 
Many real systems, however, are comprised of networks heterogeneous in nature, for example, social networks may have both agnostic and affiliative layers. 
Here, we use rank aggregation methods to identify intervention targets in multiplex networks when the structure, the dynamics, and our intervention goals are qualitatively different for each layer.
Our approach is to rank the nodes separately in each layer considering their different function and desired outcome, and then we use Borda count or Kemeny aggregation to identify a consensus ranking -- top nodes in the consensus ranking are expected to effectively balance the competing goals simultaneously among all layers.
To demonstrate the effectiveness of consensus ranking, we apply our method to a degree-based node removal procedure such that we aim to destroy the largest component in some layers, while maintaining large-scale connectivity in others.
For any multi-objective intervention, optimal targets only exist in the Pareto-sense; we, therefore, use a weighted generalization of consensus ranking to investigate the trade-off between the competing objectives. 
We use a collection of model and real networks to systematically investigate how this trade-off is affected by multiplex network structure. 
We use the copula representation of the multiplex centrality distributions to generate model multiplex networks with given rank correlations.
This allows us to separately manipulate the marginal centrality distribution of each layer and the interdependence between the layers, and to investigate the role of the two using both analytical and numerical methods.
\end{abstract}

\maketitle

\section{Introduction}

In complex networks, a small subset of nodes often has disproportionate influence on the behavior of the system~\cite{Freeman1978, Page1999, Albert2000, Newman2010}, and controlling such nodes allows us to steer the network to desired states~\cite{Motter2002, Liu2011, Fiedler2013}. For example, vaccinating a small fraction of carefully selected nodes can suppress large-scale disease outbreaks~\cite{Pastor2002}, or in a social network information is often disseminated by a small set of influencers~\cite{Kitsak2010}.
Network centrality measures -- such as degree, PageRank, or betweenness centrality  -- rank nodes based on their importance with respect to some process of interest. Therefore, high-centrality nodes provide effective targets to influence the behavior of a system: a node with high eigenvector centrality is an ideal target for vaccination~\cite{Masuda2009}, or a node in a social network with high closeness centrality is an effective influencer~\cite{Freeman1978}.
Nodes, however, typically participate in multiple networks simultaneously forming a multilayer or multiplex network~\cite{Kivela2014,Boccaletti2014,Bianconi2018}. 
For example, a person typically participates in a number of friendship and professional social networks, and these layers are often qualitatively different in nature, including affiliative and competitive interactions~\cite{Harary1953, Wasserman1994}.
Perturbing a node to influence one layer inadvertently affects the other layers that the node is active in, which can have drastic unintended consequences~\cite{Motter2002,Sornette2012,Yu2016,Lin2018}.
Therefore, it is desirable to identify targets in a multiplex network that maximize our influence in certain layers, yet minimize any unwanted impact on others. 
More generally, we consider the scenario where the structure, the dynamics, and our intervention goals are qualitatively different for each layer of a multiplex network, and we aim to find nodes that effectively balance the trade-offs between the multiple objectives.

Recently, a number of centrality measures were introduced to rank nodes in multilayer or multiplex networks~\cite{Halu2013, Estrada2014, Sole2016, Rahmede2017}. These centralities typically assume qualitatively similar layers with similar dynamics; and therefore they may provide useful intervention targets if the structure, the dynamics, and our intervention goals are all similar for each layer. 
For example, the family of ``versatility centralities'' extend common single-layer centralities to multilayer networks; as such, PageRank versatility is useful to identify influential scientists in a multiplex co-authorship network~\cite{Perc2010,Menichetti2014,DeDomenico2015a}; or airports with high betweenness versatility are prone to congestion in the multiplex air traffic network~\cite{DeDomenico2015}.
Such multiplex centralities, however, do not address the more complex, and therefore less studied, scenario where the dynamics on the layers or our objectives are qualitatively different. 
For example, social and political multiplex networks are often comprised of cooperative (friendship, collaboration, alliance) and competitive (hostility, fighting, distrust) layers~\cite{Harary1953, Wasserman1994, Maoz2010}; possible interventions might aim to reduce hostility while maintaining cooperation.
In fact, this is a practical issue that caretakers of captive nonhuman primate groups face: sometimes unusually aggressive animals are removed with the goal of reducing overall levels of conflict, and it is desirable that this removal does not negatively affect the cohesion of the group~\cite{Judge1994,Mccowan2018,Posfai2018}.
For another example, consider the recent work modeling brain activity as a two-layer multiplex network with synchronization and transport dynamics. A possible strategy to affect the behavior of the system is to target nodes that are influential in both layers~\cite{Nicosia2017}. 
Or another recent work studied a two-layer SIS model where one layer represents the spread of a disease and the other the spread of awareness of that disease~\cite{Granell2013}. 
Although both layers are characterized by SIS dynamics, a possible intervention would have different goals for the two layers: blocking the propagation of the disease and promoting awareness; and the best spreaders and the best blockers are known to have different properties~\cite{Radicchi2017}.
In these example multiplex systems, the function of each layer and/or our corresponding intervention goals are qualitatively different and ranking nodes in such networks has so far remained an uninvestigated problem. 

In this paper, we explore the use of rank aggregation methods to identify target nodes for multi-objective interventions in multiplex networks.
Instead of identifying influential nodes based on a single integrated multiplex centrality measure, we rank the nodes in each layer separately based on a centrality that is relevant to the function of the particular layer and our objectives.
 We then use rank aggregation methods to find a consensus ranking; high-ranking nodes are expected to balance all the objectives simultaneously. An expanding number of rank aggregation methods exist, and although these were originally studied by economists in the context of social choice theory~\cite{Arrow2012}, many also found applications in other disciplines, from computer science to biology~\cite{Dwork2001,Lin2010}. In this paper, we focus on two widely-used methods, Borda count and Kemeny aggregation~\cite{Kemeny1962,deBorda1781}. 
 In Sec.~\ref{sec:consensus}, we introduce consensus ranking in multiplex networks, and we demonstrate several of its properties using model and real networks. 
As an example application, in Sec.~\ref{sec:removal}, we study degree-based node removal of multiplex networks, such that we aim to destroy the largest connected component of some layers, while maintaining connectivity in the rest. In case of multiple objectives, optimality only exists in the Pareto-sense; therefore we rely on a generalized version of the Borda count algorithm that allows us to assign varying weight to the different objectives. Using the weighted Borda count, we systematically study the trade-off between the competing goals, using analytically solvable model networks and a set of real networks.

\section{Consensus ranking}\label{sec:consensus}

A multiplex network consists of $L$ layers, where each layer is a network $\mathcal G_\alpha$ ($\alpha=1,2,\ldots, L$) with the same set of nodes $\{v_1,v_2,\ldots, v_N\}$. These layers represent distinct types of interactions or relationships between the nodes, and there is no restriction on the structure of the individual layers, e.g., some layers might be directed, while others undirected, or they can be weighted or unweighted. To quantify the importance of a node $i$ in an individual layer $\alpha$, we calculate the node's centrality $c_{\alpha,i}$ as if layer $\alpha$ is in isolation. 
Nodes with high centrality are important for the functioning of that given layer, and targeting such nodes is an effective strategy to influence or monitor the behavior of that particular layer.
The type of centrality measure that works best depends on the function of the layer, for example, betweenness centrality is used for layers representing a transportation network, or eigenvector centrality is useful for layers governed by diffusion-like dynamics. However, nodes in a multiplex network do not perform a single task, but they simultaneously participate in all layers, prompting the question: How do we identify nodes that are important in all or most layers? 
This is particularly challenging if layers perform qualitatively different functions or our criteria for ranking nodes is different for each layer. To overcome these difficulties, instead of directly combining the centralities of each layer, we first determine the node ranks in each layer and then we use rank aggregation methods to identify a consensus ranking.
An expanding number of rank aggregation methods exist~\cite{Arrow2012, Dwork2001, Lin2010}, here we focus on two widely-used and intuitive methods: the Borda count (BC) and the Kemeny aggregation (KA)~\cite{deBorda1781,Kemeny1962}.

We define the rank of node $i$ in layer $\alpha$ as $r_{\alpha, i} = N - b_{\alpha, i}$, where 
\begin{equation}\label{eq:outranked}
b_{\alpha, i} = \lvert \{j\in\{1,2,\dots,N\}:\,c_{\alpha, i}>c_{\alpha, j}\}\rvert
\end{equation}
is the number of nodes that have smaller centrality than node $i$. We denote the list of node ranks in layer $\alpha$ as $r_\alpha=(r_{\alpha,1},r_{\alpha,2},\ldots,r_{\alpha,N})$. If there are no ties in the rankings (i.e., $c_{\alpha,i}\neq c_{\alpha,j}$ if $i\neq j$) $r_{\alpha}$ provides an ordinal ranking, the top rank being $1$. If ties exist, the ranks of tied nodes are set to be equal and assigned the worst value, e.g., if nodes $i$ and $j$ have equal centralities $c_{\alpha, i}=c_{\alpha, j}$ and are top ranked, we assign $r_{\alpha, i}=r_{\alpha, j}=2$ to both nodes.

The BC algorithm is perhaps the most straight-forward rank aggregation method~\cite{deBorda1781}, it works by assigning a score $b_i$ to each node $i$ equal to the average number of nodes that $i$ outranks in the layers of the multiplex:
\begin{equation}
b_i = \frac{1}{L}\sum_{\alpha=1}^L b_{\alpha, i},
\end{equation}
where $b_{\alpha, i}$ is defined in Eq.~(\ref{eq:outranked}). The BC ranking, $R_\T{BC}$, is simply a ranking of the nodes based on the scores $b_i$. The algorithm requires sorting the nodes; therefore its computational complexity is $O(N\log N)$.

The KA algorithm belongs to a larger class of rank aggregation methods that aim to identify a consensus ranking $R$ such that $R$ is at the minimum average distance from the rankings $r_\alpha$~\cite{Irurozki2014,Ceberio2015}. Formally, it is defined as
\begin{equation}\label{eq:kemcons}
R = \argmin_s \frac{1}{L}\sum_{\alpha=1}^L D(r_\alpha, s),
\end{equation}
where $D(r,s)$ is some measure of distance between two rankings $r$ and $s$, such as Hamming distance, Cayley distance, etc. 
 The KA algorithm uses the Kemeny distance~\cite{Kemeny1962}, which naturally takes into account ties:
\begin{equation}\label{eq:kemdist}
D(r,s) = \frac{1}{N(N-1)}\sum_{i<j}\lvert r(i,j)-s(i,j)\rvert,
\end{equation}
where
\begin{equation}
r(i,j) = 
\begin{cases}
	1,& \T{if } r_i < r_j, \\
	0,& \T{if } r_i = r_j, \\
	-1,& \T{if } r_i > r_j, \\
\end{cases}
\end{equation}
and $s(i,j)$ is defined similarly. Kemeny distance is normalized such that if $r$ and $s$ are the same, $D(r,s)=0$; and if $r$ and $s$ rank nodes in opposite order,  $D(r,s)=1$. If there are no ties in the rankings, the Kemeny distance is equivalent to Kendall's distance~\footnote{Note that if ties are present Kemeny distance is not a distance in the mathematical sense, as it does not satisfy the triangle inequality.}.

Unfortunately, identifying the Kemeny consensus is an NP-complete problem for $L\geq 4$~\cite{Bartholdi1989}; therefore approximate methods have to be used. Over the years, a high number of such algorithms were proposed: a recent survey compared 104 algorithms and combinations of algorithms~\cite{Ali2012}. In fact, the previously introduced BC algorithm can be considered as a simple heuristic approximation of the KA problem. 
Here, we implemented a local search algorithm, which was identified as providing optimal trade-off between accuracy and run-time~\cite{Ali2012}. 
The algorithm starts from $R_\T{BC}$ and outputs a new consensus ranking $R_\T{KA}$ by finding a local minimum of Eq.~(\ref{eq:kemcons}) using a restricted set of transformations.
The computational complexity of our implementation is $O(N^2)$; we provide the details of the algorithm in Appendix~\ref{app:LS}.

\subsection{Model networks}

We now compare the two algorithms by applying them to a multiplex network model with tunable pairwise Kemeny distance between its layers. To generate these multiplex networks, we first independently create $L$ layers, and we determine the rank $r_{\alpha,i}$ of each node $i$ in each layer $\alpha$ with respect to some centrality.
 We then generate a node label sequence for each layer to ensure that the distance between each pair of layers is $D$. Here, we use the Erd\H os-R\'enyi model (ER) or the scale-free static model (SF) to generate each layer, and we use degree centrality to rank the nodes. We describe the procedure in detail in Appendix~\ref{app:model}.

By choosing parameter $D$ of the model network, we set the strength of consensus: if $D=0.5$, the layers are independent and there is no intrinsic consensus; if $D=0$, the ranking in each layer is the same and consensus is perfect. 
We compare the two algorithms by calculating the average distance of the layers from consensus:
\begin{equation}\label{eq:avgD}
D^\T{BC/KA}_\T{con} = \frac{1}{L}\sum_{\alpha=1}^L D(r_\alpha, R_\T{BC/KA}),
\end{equation}
where $R_\T{BC/KA}$ is the consensus ranking found by the $BC$ or $KA$ algorithm. Note that $D_\T{con}$ corresponds to the cost function of the KA algorithm provided by Eq.~(\ref{eq:kemcons}), and we chose $D_\T{con}$ for comparison because BC is sometimes considered as an approximation of KA~\citep{Ali2012}. The results, however, have to be interpreted with care, finding a ranking that corresponds to lower $D_\T{con}$ doesn't necessarily mean that it is better, because ultimately BC and KA define the consensus in different ways. Similarly to the fact that no single best centrality measure exists to rank the nodes, the preferred rank aggregation algorithm also depends on our particular purposes. 

\begin{figure}[b]
	\centering
	\includegraphics[width=.5\textwidth]{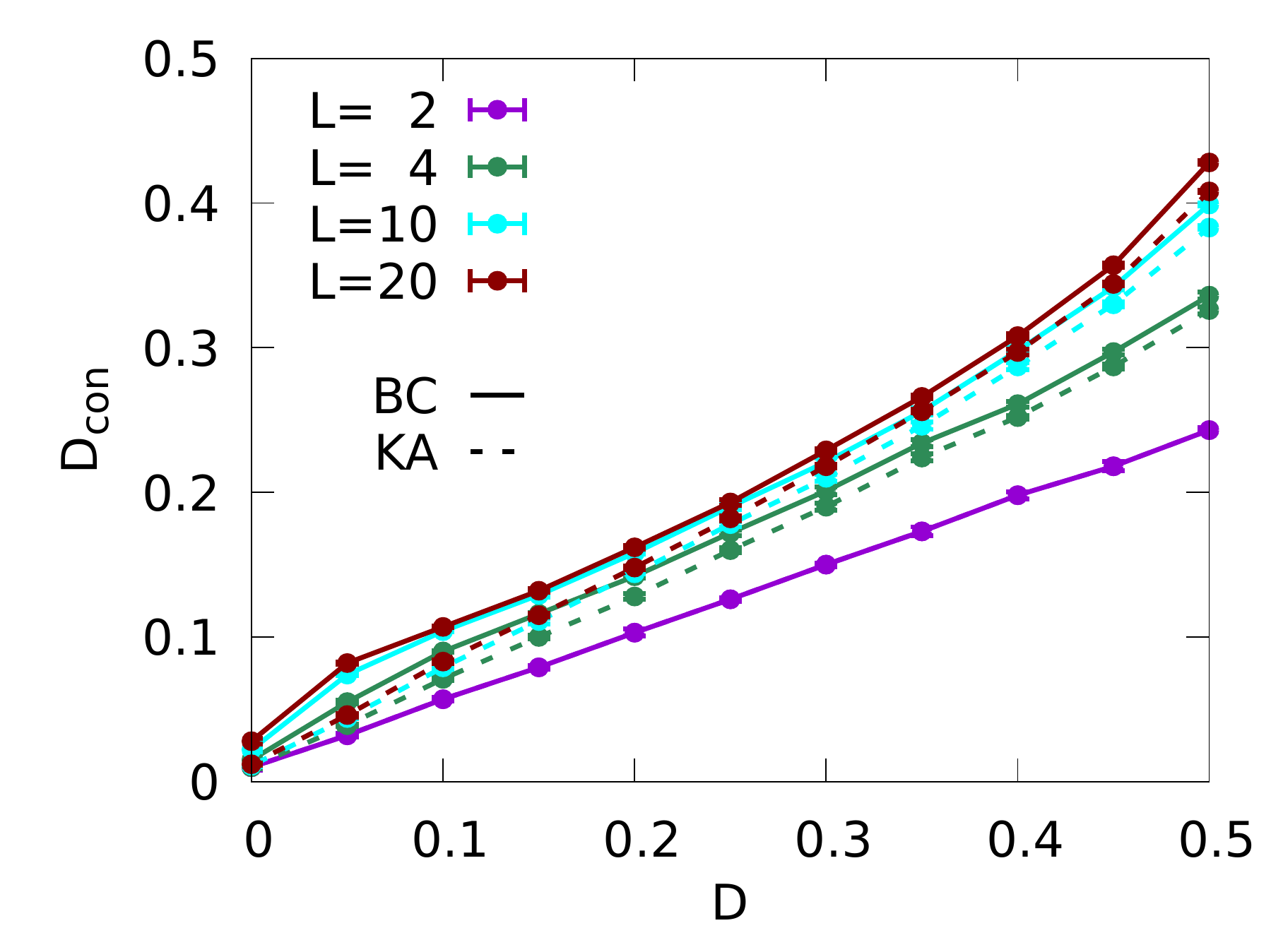}
	\caption{{ \bf Consensus ranking in model networks.} We compare the BC (solid) and KA (dashed) rankings in terms of the average distance from consensus $D_\T{con}$ as a function of number of layers $L$ and Kemeny distance between layers $D$. As expected, the KA algorithm always finds a consensus corresponding to lower $D_\T{con}$; however, the improvement is marginal. For $L=2$, both algorithms find the optimal consensus. We used ER layers with $N = 10^3$ and $c=3$; the results for SF networks are almost identical. Data points are an average of 10 independent realizations, errorbars indicate the standard error of the mean.}
	\label{fig:modelcons}
\end{figure}

Figure~\ref{fig:modelcons} shows $D_\T{con}$ as a function of the distance between layers $D$. Since our implementation of the KA algorithm works by improving the solution of the BC algorithm, we always find that $D^\T{KA}_\T{con}\leq D^\T{BC}_\T{con}$ (the equality holds for $L=2$, where both algorithms find the same consensus, corresponding to the global minimum of $D_\T{con}$). However, this improvement is marginal in terms of $D_\T{con}$.  
If strong consensus exists ($D\approx 0$) both algorithms closely approximate this consensus. Even if the layers are independent ($D\approx 0.5$), we are able to find a consensus with $D_\T{con}<0.5$, especially for multiplexes with only a few layers. For the latter case, the consensus does not capture an intrinsic property of the multiplex network, rather it identifies nodes that by chance have high rank in many layers; and therefore potentially provide low-cost targets for influencing multiple layers simultaneously.

Note that Fig.~\ref{fig:modelcons} shows results for ER layers, we found that  using the SF layers produces almost indistinguishable results. In fact, the performance of the algorithms only depends on the rankings $r_\alpha$, and not on the details of each layer $\alpha$. The only property of $r_\alpha$ that is specific to ER or SF layers is the number of ties in the ranking, and our results indicate that the performance of the BC and KA algorithms in terms of $D_\T{con}$ are insensitive to this.

\subsection{Real networks}\label{sec:realcons}

The BC and KA algorithms allow us to identify and analyze consensus ranking in real systems represented by multiplex networks. Here, we investigate three examples:
\begin{enumerate}
\item Airline network. A multiplex network with 5 layers representing the United States air traffic network in 2013. 
The nodes are airports and links indicate direct flights between them, and the layers correspond to the 5 largest carriers~\footnote{The airline network represents the five largest US carriers in terms of number of unique flights. It is constructed based on the publicly available database of the USA Department of Transportation (\url{https://www.transtats.bts.gov/Fields.asp?Table_ID=259}), all domestic flights labeled as ``scheduled passenger service'' are included.}.
\item Primate social network. A multiplex network with 4 layers containing interaction data from one week of observations of a captive rhesus macaque troupe. The layers represent different interactions between animal pairs, including conflict, signaling of subordination, grooming, and huddling~\cite{Mccowan2011, Beisner2015}.
\item Human social network. A multiplex network with 5 layers, where nodes are members of the Department of Computer Science at the Aarhus University and the links indicate various social relationships: Facebook friendship, spending leisure time together, working together, co-authorship, and regularly having lunch together~\cite{Magnani2013}.
\end{enumerate}
Table~\ref{tab:realprop} summarizes some basic properties of these networks. The airline network has a clear heavy-tailed degree distribution with hubs that have significantly more connections than average nodes.
 It is not possible to be as definite about the social networks due to their small size; there are, however, nodes that are connected to a significant fraction of the network, for example, in the fifth layer of the human social network the largest hub is connected to almost half of the other nodes. 
The primate social network is composed of competitive interactions (conflict and subordination) and cooperative interactions (grooming and huddling) -- the existance of antagonistic and affiliative layers is a general property of social and political networks~\cite{Harary1953, Wasserman1994, Maoz2010}. For the primate network, we find that the adversarial layers are more heterogeneous than the affiliative layers. Interestingly, similar pattern was observed for human social networks obtained from an online game that allowed competition and alliances between players~\cite{Szell2010}.

\begin{figure}[t]
	\centering
	\includegraphics[width=1.\textwidth]{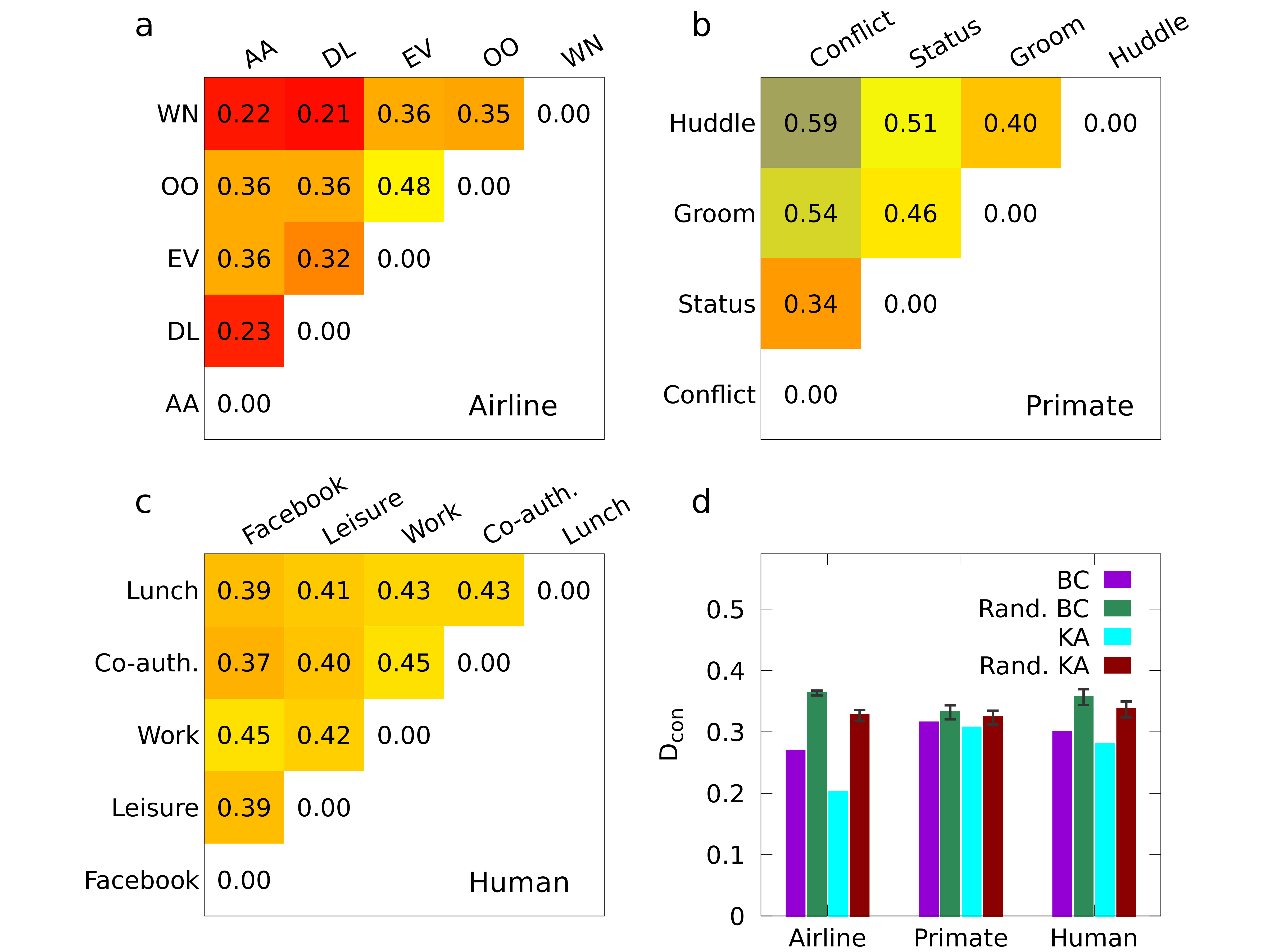}
	\caption{{ \bf Consensus ranking in real networks.} (a-c) The pairwise Kemeny distance between layers in the airline network, primate social network, and human social network. If layers are independent, $D=0.5$ (yellow); if rankings of layers are identical, $D=0$ (red). (d)~Average distance from consensus $D_\T{con}$ for the BC (purple) and KA (blue) rankings, compared to a randomized null-model obtained by shuffling node labels in each layer (green and red). $D_\T{con}$ values reported for randomizations are an average of 1000 independent realizations, and errorbars indicate the standard deviation.}
	\label{fig:realcons}
\end{figure}

When ranking nodes in a network, we chose a node centrality depending on the function of the system and our goals. Here, for illustration purposes, we calculate the node rankings for each network based on degree centrality. Figure~\ref{fig:realcons}(a-c) shows the pairwise Kemeny distances between layers for each network. Typically, we find $D<0.5$, indicating positively correlated rankings; the correlation is the strongest in the airline network, while rankings in the social networks are less aligned with each other. 
We observe an interesting pattern in the primate social networks: the distance is low between affiliative layers (grooming and huddling), and also low between the interactions involving social hierarchy (conflict and signaling); but the rankings corresponding to competitive layers are independent from rankings of affiliative layers.

We also identify the consensus ranking for each network using the BC and the KA algorithms. Figure~\ref{fig:realcons}d shows the average distance of the layers from consensus, $D_\T{con}$, for the original networks and their randomized counterparts, where we shuffled the node labels in each layer, eliminating any correlation between the rankings. For all multiplexes, we find  stronger consensus in the real instances than in their randomized versions: the largest difference was observed in the airline network, while smallest in the primate network. As expected, KA always finds a consensus with a lower $D_\T{con}$ than BC; the difference, however, is marginal with the exception of the airline network.

Overall, we found for both model and real multiplexes that the BC and KA consensus rankings are similar in terms of $D_\T{con}$, while the BC algorithm is faster and extremely simple to implement. Both algorithms effectively identify the consensus if the layers are strongly correlated, and even if the layers of a multiplex are independent, we can identify a consensus ranking that provides low-cost targets for simultaneous intervention on multiple layers. So far, we only compared the consensus rankings in terms of $D_\T{con}$, in the next section we demonstrate how consensus ranking can identify  targets for a simple degree-based node removal model with multiple objectives.

\begin{table}[b]
\centering
\begin{tabular}{l|c|c|c|c|c}
& $\ \ N\ \ $ & $\ \ L\ \ $ & $\avg{k}$ & $\kappa=\nicefrac{\langle{k^2\rangle}}{\avg{k}}$ & $k_\T{max}$\\
\hline\hline
Airline        & 575 & 5 & \makecell{4.63, 2.95, 9.02,\\5.99, 4.11} &  \makecell{28.7, 39.1, 37.2,\\31.9, 28.3} & \makecell{109, 126, 125,\\95, 66} \\
\hline
Primate Soc.   & 100 & 4 & \makecell{8.62, 4.90,\\4.32, 4.98} & \makecell{12.1, 8.9,\\5.8, 6.3} & \makecell{35, 29,\\13, 12}\\
\hline
Human Soc.     & 61 & 5 & \makecell{6.33, 4.07, 0.69,\\2.89, 6.36} & \makecell{7.8, 9.5, 2.3,\\5.7, 10.8} & \makecell{15, 15, 5,\\14, 27}
\end{tabular}
\caption{\label{tab:realprop} {\bf Properties of the example multiplex networks.} We provide the number of nodes ($N$) and layers ($L$), the average degree of each layer ($\avg k$); and we quantify the degree heterogeneity of each layer with   $\kappa = \avg{k^2}/\avg{k}$ (for Poisson degree distribution $\kappa = 1+\avg k$ and for scale-free networks $\kappa\rightarrow \infty$ as $N\rightarrow \infty$) and with the maximum degree $k_\T{max}$.}
\end{table}

\section{Multi-objective degree-targeted attack}\label{sec:removal}

In this section, we explore an example of using consensus ranking to identify effective targets for multi-objective interventions in multiplex networks. A classic result of network science is that complex networks with heavy-tailed degree distributions are robust against random node removal; while targeted removal of high-degree hubs rapidly breaks down their large-scale connectivity~\cite{Albert2000}. Complex networks, however, rarely exist in isolation, nodes typically participate in multiple networks simultaneously. Therefore, removing a node from a multiplex network to reduce connectivity in one layer, might also remove connections from other layers that otherwise we would like to preserve. More specifically, here, we consider the problem of removing nodes from a multiplex network with $L$ layers such that for a set of layers $\mathcal L_\T D$ we aim to reduce the size of the largest component, while we want to keep the rest of the layers  $\mathcal L_\T K$ intact. We rank the nodes in each layer based on degree centrality, but to reflect our different objectives, we reverse the rankings for the layers that we keep intact:
\begin{equation}\label{eq:DKrank}
\begin{split}
r^\T D_{\alpha, i} &= N - \lvert \{j\in\{1,\dots,N\}:\,k_{\alpha, i}>k_{\alpha, j}\}\rvert,\\
r^\T K_{\alpha, i} &= N - \lvert \{j\in\{1,\dots,N\}:\,k_{\alpha, i}<k_{\alpha, j}\}\rvert,
\end{split}
\end{equation}
where $r^\T D_{\alpha, i}$ and $r^\T K_{\alpha, i}$ are the ranks of node $i$ in layers that we want to destroy and keep, respectively; and $k_{\alpha, i}$ is the degree of node $i$ in layer $\alpha$. Given a multiplex network, we calculate these ranks for each layer, and we use the BC and KA algorithms to identify consensus rankings. We then iteratively remove the nodes based on this consensus ranking starting with the highest ranked nodes.

\begin{figure}[t]
	\centering
	\includegraphics[width=1.\textwidth]{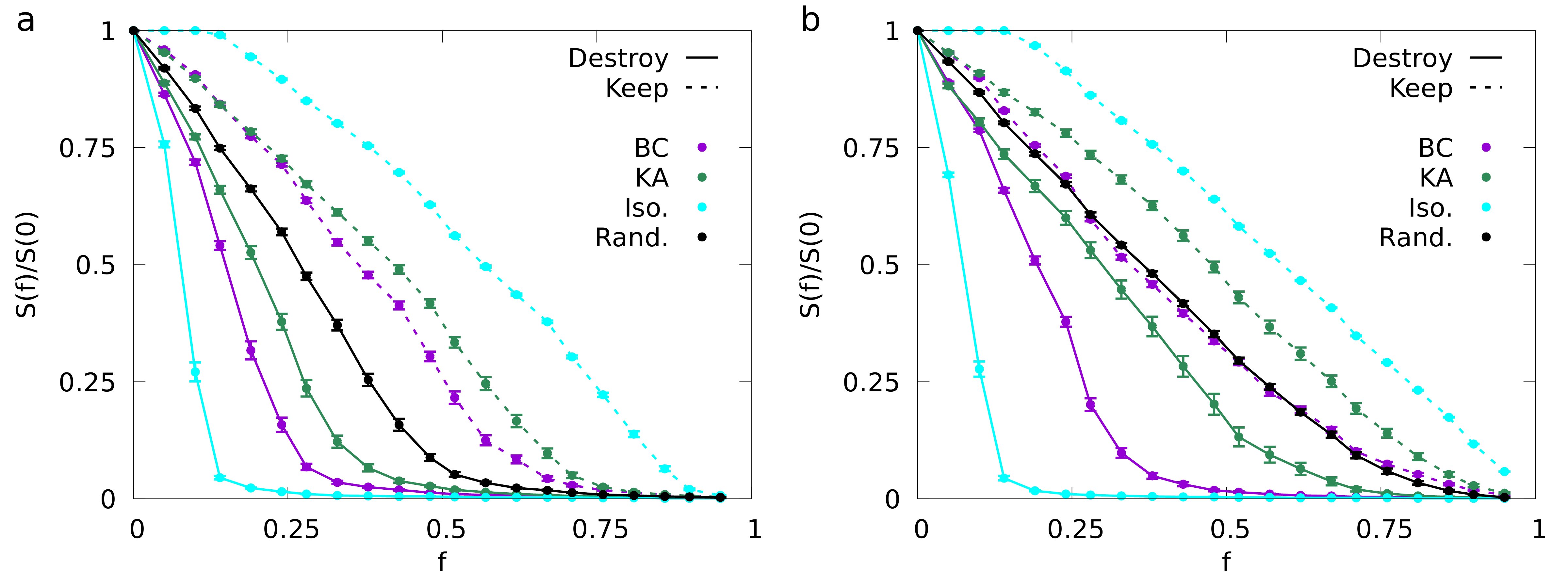}
	\caption{{ \bf Consensus-based removal.} We compare BC (purple) and KA (green) consensus-based removal to random removal (black) and removal as if layers are in isolation (blue). We show one of the layers that we destroy (solid line) and one of the layers that we preserve (dashed line). We use model networks with parameters (a)~$N=10^3$, $L=4$, $D=0.4$, ER layers  with $c=3$; and (b)~$N=10^3$, $L=4$, $D=0.3$, SF layers with $c=3$, $\gamma=2.5$. Data points are an average of 100 independent realizations, errorbars represent the standard error of the mean.}
	\label{fig:modellcc}
\end{figure}

Figure~\ref{fig:modellcc} shows the relative size of the largest connected component $S(f)$ as a function of the fraction of nodes removed $f$ for model multiplex networks with $L=4$ ER or SF layers (for details about the model networks see Appendix~\ref{app:model}). For the ER example, we find that both the BC and KA consensus-based removal reduces $S$ faster for the layers we aim to destroy, and slower for layers we aim to preserve than random node removal, but BC preforms better at reducing $S$, while KA is better at preserving connectivity. Similarly for the SF example, BC destroys the targeted layers faster than KA, but doesn't keep the rest of the layers intact more than random removal would. 
In either case, we are not as effective as if the layers would be in isolation. 

Comparing the BC and KA rankings in Fig.~\ref{fig:modellcc}, we find that neither method is objectively better than the other, instead they provide a different trade-off between the two objectives: BC preforms better at destroying layers, while KA does better at keeping layers intact. In fact, this is a pattern that we widely observed varying the parameters of the multiplex model networks. Therefore, from hereon in this section, we will focus on the BC ranking and we will study the trade-off by introducing a weighted version of BC. 
Further reasons for focusing on the BC algorithm are that it is much faster than the KA and the simplicity of the BC algorithm allows us to analytically solve $S(f)$ for the model networks.

The original BC algorithm provides one possible trade-off between the competing objectives of destroying certain layers, while keeping others intact. To explore other possible trade-offs, we introduce a weighted Borda count (wBC) algorithm that allows us to assign varying preference to the different objectives. The weighted Borda score of node $i$ is defined as
\begin{equation}\label{eq:wbordascore}
b_i = \frac{1}{L}\sum_{\alpha=1}^L w_\alpha b_{\alpha, i},
\end{equation}
where $w_\alpha = w$, if we destroy layer $\alpha$ ($\alpha \in \mathcal L_\T D$); $w_\alpha = 1-w$, if we keep layer $\alpha$ intact ($\alpha \in \mathcal L_\T K$); and $w\in[0,1]$. The choice $w=0.5$ corresponds to the unweighted Borda score (with a multiplier of 1/2); if $w=1$, we only care about the layers we aim to destroy; and if $w=0$, we only care about the layers we aim to keep intact. In the following, we  first derive an analytical solution for $S(f)$, and then we systematically investigate how multiplex network structure  affects the trade-off between the different objectives using model and real networks.

\subsection{Analytical solution}\label{sec:analytic}

To analytically solve the size of the giant connected component (GCC)  for consensus-based removal, we first calculate the size of the GCC for a general degree-based removal strategy (DBS) on a single layer network, and then we map the consensus-based process to a degree-based one for each layer. By DBS we mean a node removal process where the probability of removing a node only depends on its degree, i.e., the probability of removing node $i$ is $f(k_i)$. Let $s$ be the probability that a randomly selected link leads to the GCC. Assuming local tree-like structure and uncorrelated networks, we calculate $s$ using the self-consistent equation
\begin{equation}\label{eq:s}
1- s	= \sum\limits_{k=1}^\infty \frac{k}{c}\left(1-f(k)\right)p(k)(1-s)^{k-1} + \sum\limits_{k=1}^\infty \frac{k}{c} f(k)p(k),
\end{equation}
where $p(k)$ is the degree distribution and $c$ is the average degree of the network. The first term on the right hand side is the probability that following a random link leads to a node that is not removed and not part of the GCC, and the second term is the probability that the node is removed. Once $s$ is determined, we obtain the relative size of the GCC using the equation
\begin{equation}\label{eq:S}
1- S	= \sum\limits_{k=0}^\infty \left(1-f(k)\right)p(k)(1-s)^{k} + \sum\limits_{k=0}^\infty f(k)p(k).
\end{equation}
Several node removal processes can be described as a DBS. For example, the choice $f(k)\equiv f$ leads to simple random node removal.
Generally the best DBS to destroy a single-layer network removes nodes starting with the highest degree. Formally we express this as
\begin{equation}\label{eq:f_hideg}
f(k) = 
\begin{cases}
	1,& \T{if } k > K, \\
	\frac{f-\sum_{k>K}p(k)}{p(K)},& \T{if } k = K, \\
	0,& \T{if } k< K, \\
\end{cases}
\end{equation}
meaning that we remove all nodes with degree higher than $K$ and a certain fraction of nodes with degree $K$, where $K=\inf\{k :1-f\leq P(k)\}$ and $P(k)$ is the CDF of the degree distribution. Also, the most effective DBS to keep a single-layer network intact is to remove nodes with the lowest degree first:
\begin{equation}\label{eq:f_lodeg}
f(k) = 
\begin{cases}
	0,& \T{if } k > K, \\
	\frac{f-\sum_{k<K}p(k)}{p(K)},& \T{if } k = K, \\
	1,& \T{if } k< K, \\
\end{cases}
\end{equation}
where $K=\inf\{k :f\leq P(k)\}$.

To map the consensus-based node removal to a DBS, we first calculate the Borda score of a node with multiplex degree $(k_1,\ldots,k_L)$ as
\begin{equation}\label{eq:k_to_b}
b(k_1,\ldots,k_L) = \frac{w}{L}\sum_{\alpha\in \mathcal L_\T D}P_\alpha(k_{\alpha}-1) + \frac{1-w}{L}\sum_{\alpha\in \mathcal L_\T K}\left(1-P_\alpha(k_{\alpha})\right),
\end{equation}
where $P_\alpha(k)$ is the CDF of the degree distribution of layer $\alpha$. Next, we calculate the Borda score distribution, i.e., the probability that a randomly selected node of the multiplex has Borda score $b$:
\begin{equation}\label{eq:bdist}
p_\T B(b) = \sum\limits_{k_1,\ldots,k_L=0}^\infty p(k_1,\ldots,k_L)\delta\left(b-b(k_1,\ldots,k_L)\right),
\end{equation}
where $p(k_1,\ldots,k_L)$ is the multiplex degree distribution.
When removing $f$ fraction of nodes based on consensus ranking, the probability of removing a node with Borda score $b$ is
\begin{equation}\label{eq:f_b}
f(b) = 
\begin{cases}
	1,& \T{if } b > B, \\
	\frac{f-\sum_{b>B} p_\T B(b)}{p_\T B(B)},& \T{if } b = B, \\
	0,& \T{if } b< B,
\end{cases}
\end{equation}
 where $B = \inf\{b: 1-f\leq P_\T B(b)\}$ and $P_\T B(b)$ is the CDF corresponding to $p_\T B(b)$ provided by Eq.~(\ref{eq:bdist}). For each layer $\alpha$, the consensus-based removal can be formulated as an effective DBS with
\begin{equation}\label{eq:f_kborda}
f_\alpha(k_\alpha) = \frac{\sum_{\{k_\beta:\beta\neq\alpha\}} p(k_1,\ldots,k_L)f(b(k_1,\ldots,k_L))}{\sum_{\{k_\beta:\beta\neq\alpha\}} p(k_1,\ldots,k_L)},
\end{equation}
where $p(k_1,\ldots,k_L)$ is the joint multiplex degree distribution, and the summations are over the degrees in all layers, except $\alpha$.

\begin{figure}[t]
	\centering
	\includegraphics[width=1.\textwidth]{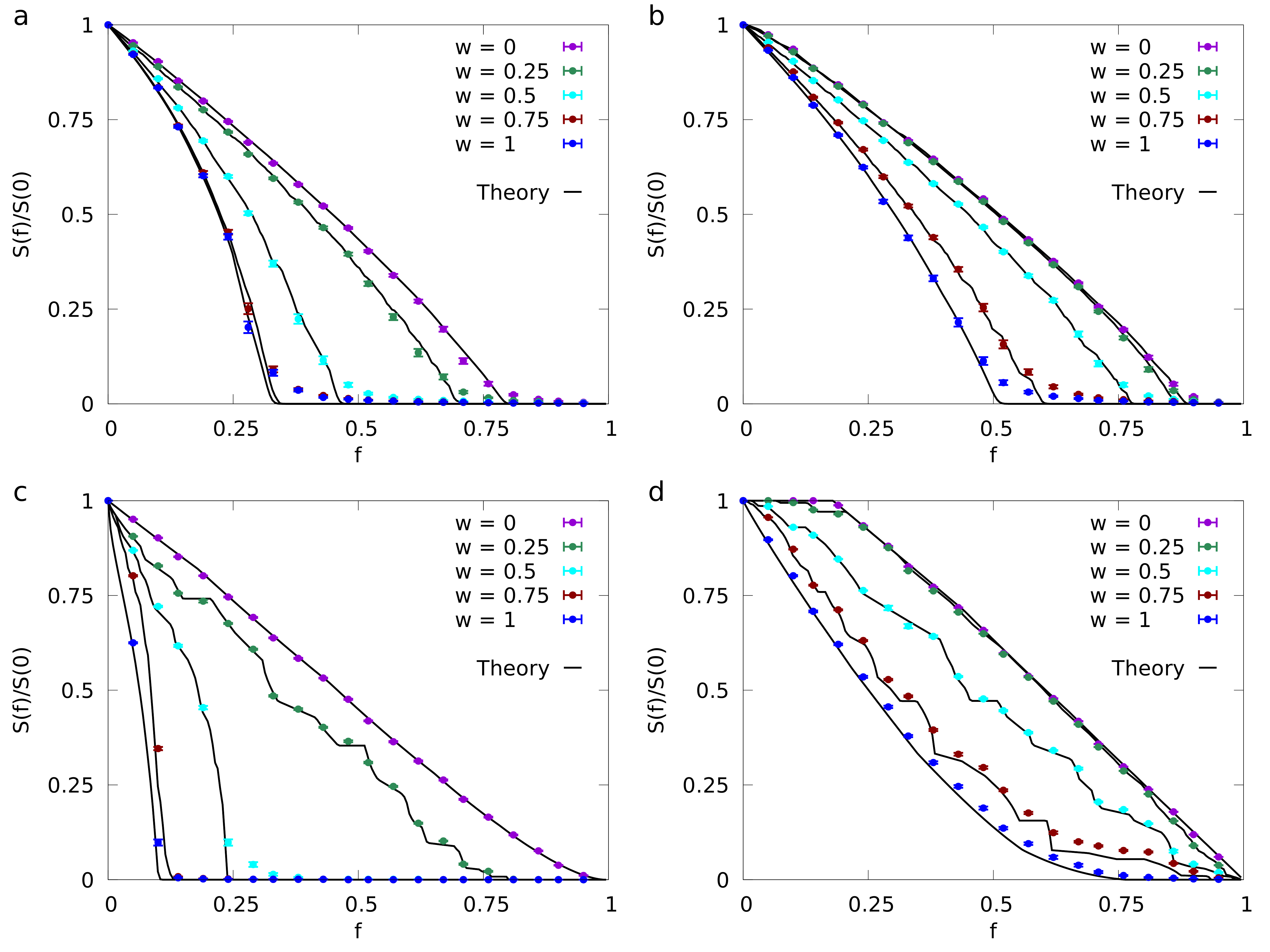}
	\caption{{ \bf Comparing theory and simulations.} We compare the relative size of the largest connected component as a function of $f$ obtained from simulations (markers) and numerical evaluation of Eqs.~(\ref{eq:S}) and (\ref{eq:f_kborda}) (continuous lines), overall finding excellent agreement for both (a,c)~layers that we destroy and (b,d)~layers that we keep intact. The theory reveals that $S$ depends on $f$ in a highly non-trivial way.  (a-b)~Multiplex network with parameters $N=10^3$, $L=4$, $D=0.4$, ER layers  with $c=3$; and (c-d)~$N=10^4$, $L=2$, $D=0.4$, SF layers with $c=3$, $\gamma=2.5$. Data points are an average of 100 independent realizations, errorbars represent the standard error of the mean. }
	\label{fig:theory-sim}
\end{figure}

Substituting Eq.~(\ref{eq:f_kborda}) into Eqs.~(\ref{eq:s}) and (\ref{eq:S}) provides $S_\alpha(f)$, the relative size of the GCC of layer $\alpha$ for consensus-based node removal. We numerically evaluate these equations for a class of multiplex model networks where the layers are either ER or SF networks, for SF networks we use the degree distribution provided in Eq.~(\ref{eq:SFdegdist}). 
Figure~\ref{fig:theory-sim} compares the analytical solution of $S_\alpha(f)$ to simulations, showing excellent agreement. We find that there is a non-trivial connection between $S_\alpha(f)$ and $f$ as the node removal process switches back-and-forth between destroying and preserving layers. 
Note that the methods that we used to numerical evaluate the necessary equations become intractable for increasing number of layers; therefore for the numerical solutions we restrict ourselves to $L=4$ ER and $L=2$ SF layers.

\subsection{Model networks.}

The $w$ parameter of the wBC algorithm allows us to assign different level of importance to the two competing objectives: $w\approx 1$ favors breaking down layers, while $w\approx 0$ focuses on keeping layers intact. To systematically investigate this trade-off, we introduce the coefficients
\begin{equation}\label{eq:O}
\begin{split}
O_\T D &= \frac{1}{\lvert \mathcal L_\T D \rvert}\sum\limits_{\alpha\in\mathcal L_\T{D}}\frac{S_\alpha-S^\T{min}_\alpha}{S^\T{max}_\alpha-S^\T{min}_\alpha},\\
O_\T K &= \frac{1}{\lvert \mathcal L_\T K \rvert}\sum\limits_{\alpha\in\mathcal L_\T{K}}\frac{S^\T{max}_\alpha-S_\alpha}{S^\T{max}_\alpha-S^\T{min}_\alpha},
\end{split}
\end{equation}
where $S_\alpha$ is the size of the GCC in layer $\alpha$ after removing $f$ fraction of its nodes using the wBC algorithm; $S^\T{min}_\alpha$ ($S^\T{max}_\alpha$) is the size of the GCC after removing $f$ fraction of its highest (lowest) degree nodes, i.e., it is the best we could do if layer $\alpha$ was in isolation. The coefficients $O_\T{D}$ and $O_\T{K}\in[0,1]$ measure how well the objectives are achieved relative to the case when we remove nodes from each layer independently. Analytical solution of Eq.~(\ref{eq:O}) is obtained by first calulcating $S^\T{min}_\alpha$, $S^\T{max}_\alpha$, and $S_\alpha$ using Eqs.~(\ref{eq:f_hideg}), (\ref{eq:f_lodeg}), and (\ref{eq:f_kborda}).

\begin{figure}[t]
	\centering
	\includegraphics[width=1.\textwidth]{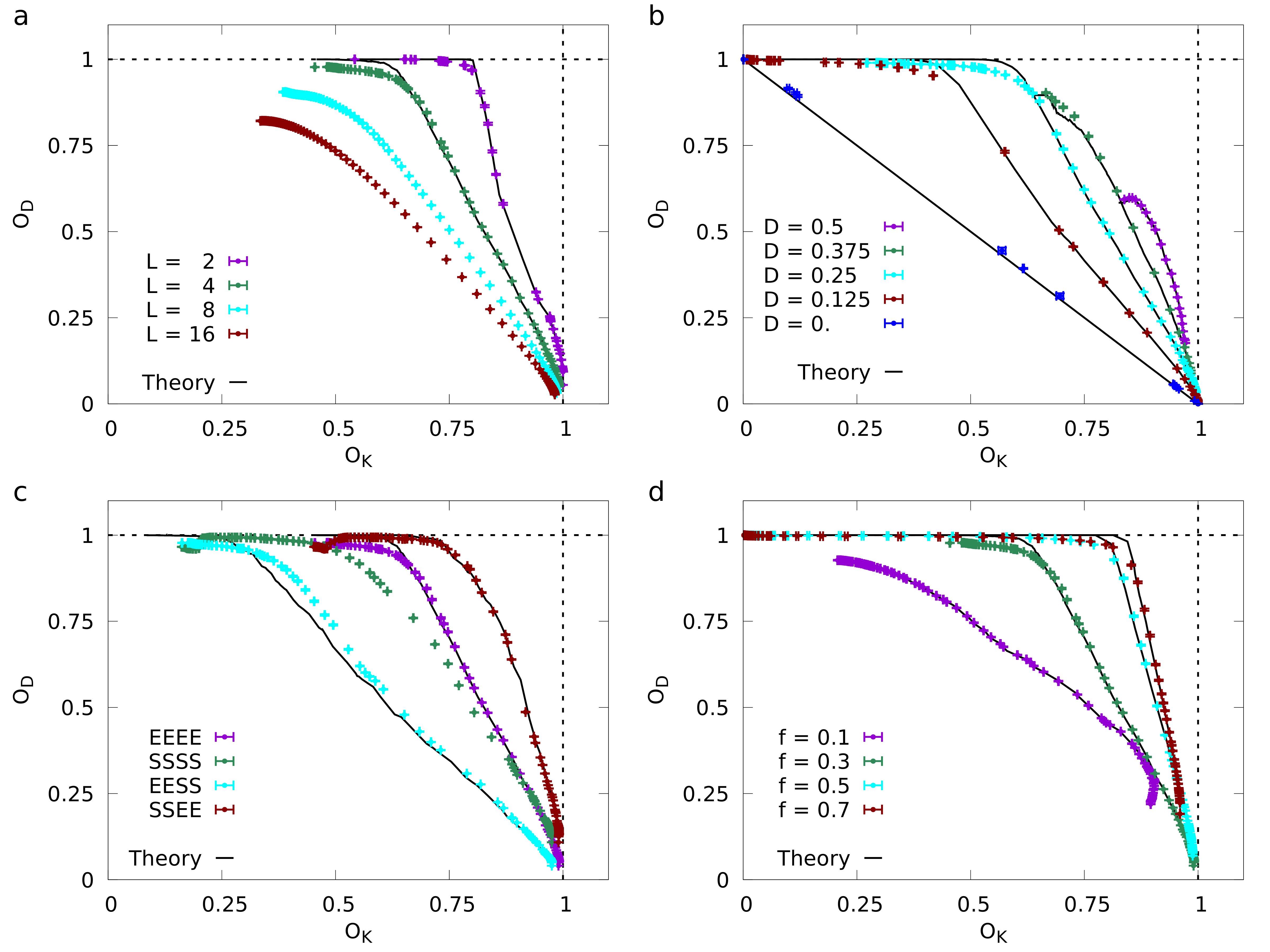}
	\caption{{ \bf Trade-off curve of model networks.} We calculate the $O_\T D$-$O_\T K$ trade-off for model networks with tunable Kemeny distance between layer pairs when we aim to destroy half of the layers, and keep the rest intact. For simulations we used parameters $N=10^4$, $L=4$, $c=3$, $D=0.3$, and $f=0.3$, unless otherwise noted on the figure; all layers are ER networks except for (c), where some layers are SF with degree exponent $\gamma=2.5$. The trade-off curves are shown varying (a)~number of layers~$L$, (b)~Kemeny distance $D$, (c)~degree heterogeneity of layers, and (d)~fraction of nodes removed $f$. Data points are an average of 1000 independent simulations, errorbars along both axis represent the standard error of the mean; continuous lines are obtained by evaluating Eq.~(\ref{eq:O}) whenever numerically tractable.}
	\label{fig:modeltradeoff}
\end{figure}

Figure~\ref{fig:modeltradeoff} shows the trade-off curve between $O_\T{D}$ and $O_\T{K}$ for model multiplex networks with $L$ layers, where we aim to destroy half of the layers, while keeping the other half intact. If there would be no trade-off, the curve would be a single point at $(O_\T{K},O_\T{D})=(1,1)$; if all layers would be identical, the trade-off curve would be the diagonal line $O_\T{D}=1-O_\T{K}$. 



Figure~\ref{fig:modeltradeoff}a shows that increasing the number of layers affects the trade-off in two distinct ways: (i)~As $L$ increases the number of conflicting objectives also increase, and the trade-off becomes more severe approaching the diagonal line. (ii)~For $L\geq4$, we aim to destroy multiple layers simultaneously, which cannot be done as efficiently as if they where in isolation; therefore even if $w=1$, $O_\T D$ remains less than 1. 
Increasing $L$ has a similar, but weaker effect on $O_\T K$, since destroying a network is more difficult than keeping it intact. 
Changing the pairwise Kemeny distance $D$ between the layers (Fig.~\ref{fig:modeltradeoff}b), we find that strong consensus ($D\approx 0$) leads to strong trade-off between $O_\T D$ and $O_\T K$, but allows efficient destruction ($\max_w O_\T D\approx 1$) or preservation ($\max_w O_\T K\approx 1$) of layers. 
Figure~\ref{fig:modeltradeoff}c shows the effect of degree heterogeneity, we find that destroying SF layers and preserving ER layers entails the least amount of trade-off, while destroying SF and preserving ER is the most difficult. Finally, Fig.~\ref{fig:modeltradeoff}d shows that the trade-off is the most pronounced at the initial stages of node removal (small $f$), and the competing objectives are less restrictive in case of large-scale removals (large $f$).

\subsection{Real networks.}

In the previous section, we used model multiplex networks to understand how basic network properties affect the $O_\T D$-$O_\T K$ trade-off. Real networks, however, have more complex structure. To investigate their effect, we use the three example datasets introduced in Sec.~\ref{sec:realcons}: the human social network, the primate social network, and the airline network. We assign half of the layers in each multiplex to be destroyed and the rest to be kept intact, and we calculate the $O_\T D$-$O_\T K$ trade-off after removing a fraction of the nodes ($f=0.4$ for the airline network and the human social network, and $f=0.55$ for the primate social network). We then compare the trade-off curves to the following randomized null models:
\begin{enumerate}
\item Full randomization (FR): We randomly rewire all links, in effect replacing each layer with an ER network with the same number of nodes and links.
\item Node label randomization (NLR): We shuffle the node labels in each layer, removing any correlation or consensus between layers, but leaving the structure of each layer unchanged.
\item Degree preserved randomization (DPR): We rewire links such that the multiplex degree of each node is unchanged. This randomization, therefore, preserves the pairwise Kemeny distances between the layers, and removes all structure within the layers beyond their degree sequence.
\item Node label and degree preserved randomization (NLR+DPR): Combining NLR and DPR removes correlations both within layers and between layers, and only preserves the degree distributions of the individual layers.
\end{enumerate}

\begin{figure}[t]
	\centering
	\includegraphics[width=1.\textwidth]{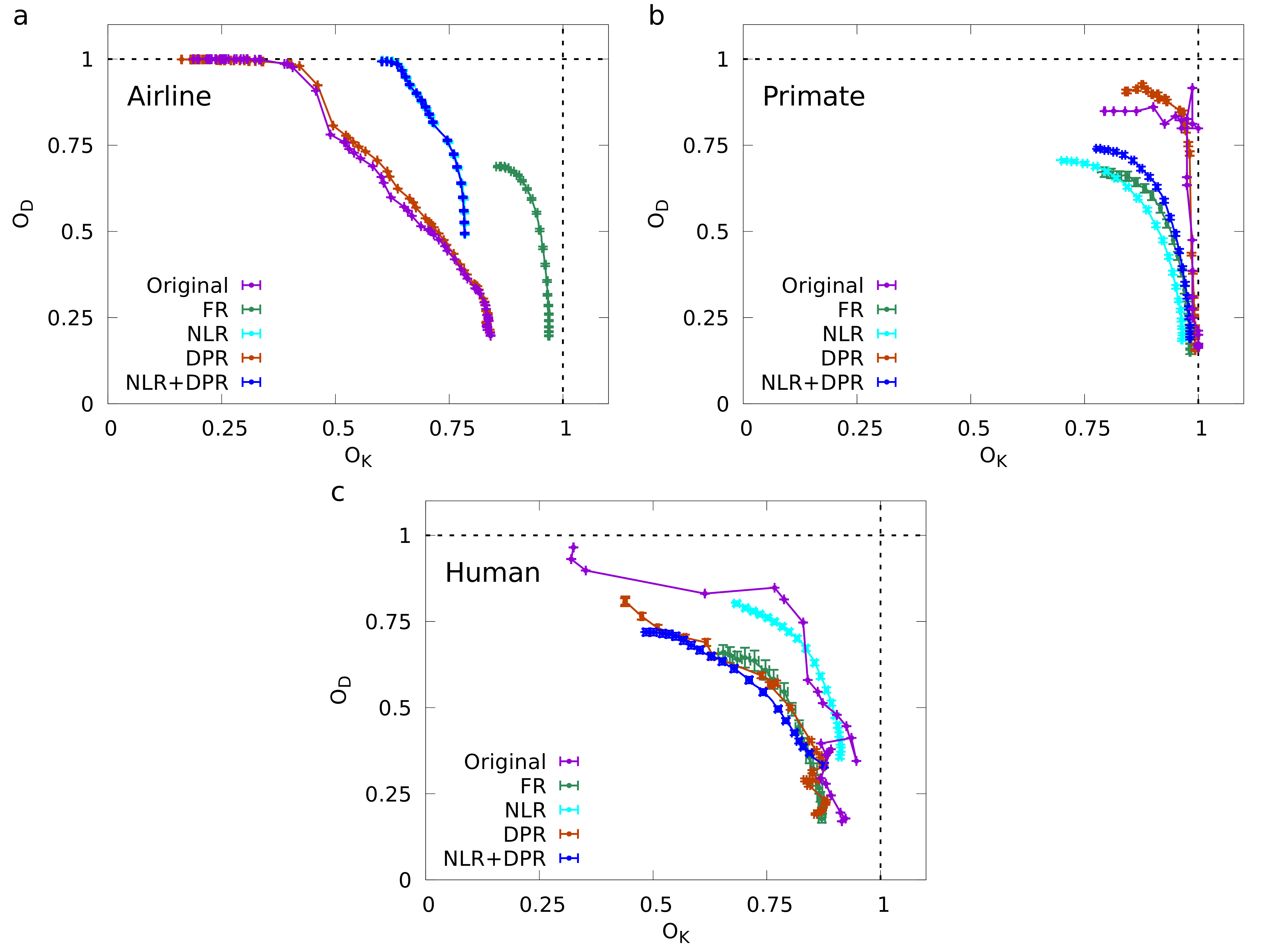}
	\caption{{ \bf Trade-off curve of real networks.} The $O_\T D$-$O_\T K$ trade-off curve is shown for the original networks (purple), full (FR, green), node label (NLR, light blue), and degree preserving randomization (DPR, red), and combining node label and degree preserving randomization (NLR+DPR, dark blue). We remove $f$ fraction of nodes, aiming to destroy layers $\mathcal L_\T D$ , while keeping layers $\mathcal L_\T K$ intact. (a)~Airline network: $f=0.4$, $\mathcal L_\T D = \{\T{AA}, \T{DL}\}$, $\mathcal L_\T K = \{\T{EV}, \T{OO}, \T{WN}\}$; (b)~Primate social network: $f=0.55$, $\mathcal L_\T D = \{\T{Conflict}, \T{Signal}\}$, $\mathcal L_\T K = \{\T{Groom}, \T{Huddle}\}$; (c)~Human social network: $f=0.4$, $\mathcal L_\T D = \{\T{Facebook}, \T{Leisure}\}$, $\mathcal L_\T K = \{\T{Work}, \T{Co-auth.}, \T{Lunch}\}$. Data points are an average of 1000 independent realizations, errorbars along both axis represent the standard error of the mean.}
	\label{fig:realtradeoff}
\end{figure}

Figure~\ref{fig:realtradeoff} provides the trade-off curves for the three example networks, each network showing a distinct behavior. 
For the airline network, we aim to destroy two layers and keep the other three intact. In Sec.~\ref{sec:realcons}, we showed that the airline network has  heterogeneous degree distributions and strong consensus between its layers. For model networks, we found that strong consensus leads to significant $O_\T D$-$O_\T K$ trade-off, and indeed, we find that for the airline network the trade-off curve is close to the diagonal.  We find that the DPR trade-off curve is almost indistinguishable from the original, while the trade-off for FR, NLR, and NLR+DPR is significantly weaker, meaning that both the inter-layer correlations and the degree distributions are necessary to explain the strong trade-off.  
We also find that $\max_w O_\T D=1$ for the original and NLR curve, but not for FR, this indicates that the simultaneous destruction of the two layers is aided by the presence of hubs and is uneffected by the inter-layer correlations.

In the case of the primate social network, we aim to destroy the two layers that are related to competition (conflict and signaling) and to preserve the affliative layers (grooming and huddling). In fact, it is common practice for the caretakers of captive primate groups to remove unusually aggressive individuals with to goal of reducing overall levels of conflict; and it is desirable that this removal doesn't negatively affect the cohesion of the group~\cite{Judge1994,Mccowan2018,Posfai2018}.
Figure~\ref{fig:realtradeoff}b shows the $O_\T D$-$O_\T K$ curves for the primate network, and we find a very weak trade-off. This is largely explained by the inter-layer correlations, which 
in contrast to the airline network, reduce the trade-off. To understand this recall Fig.~\ref{fig:realcons}b, where we showed that Kemeny distance is small between the conflict-signaling and the grooming-huddling layer pairs, while there is no or even negative correlation between the competitive and affiliative layers. This particular structure allows us to simultaneously disrupt the competitive layers without affecting the affiliative layers. Furthermore, the competitive layers have more heterogeneous degree distributions than the affiliative layers (Table~\ref{tab:realprop}), a property also seen in human social networks~\cite{Szell2010}. Using model networks, we showed that destroying heterogenous layers and keeping homogeneous layers intact reduces the trade-off (Fig.~\ref{fig:modeltradeoff}c). Indeed, comparing the FR and DPR+NLR curves shows that the degree distribution of the layers contributes to the weak trade-off, albeit less than the inter-layer correlations.
 
 
Finally, for the human social network, our goal is to reduce connectivity in two layers and keep the other three intact. For both the airline network and the primate social network, we found that the inter-layer correlations significantly affect the $O_\T D$-$O_\T K$ trade-off; interestingly, in case of the human social network inter-layer correlations have little effect. The original trade-off curve is best approximated by the NLR, which removes correlations between layers, but preserves all structure within each layer; and all other randomizations show stronger trade-off than the original. Therefore, we conclude that the internal structure of the layers beyond the degree distribution, such as community structure, is what reduces trade-off.

\section{Conclusion}

In this paper, we explored the use of consensus rankings to identify effective targets for multi-objective interventions in multiplex networks. Our strategy is to calculate a centrality for each layer that is relevant to its specific function and our specific objectives, rank the nodes based on these centralities, and then using rank aggregation methods identify a consensus ranking. As an example process, we studied the degree-based node removal process, where we aimed to destroy the largest connected component in some layers, while keeping the other layers intact. We demonstrated that removing the nodes in order of consensus ranking effectively balances these competing objectives.

The advantage of our method is that it is agnostic to the specific properties of the layers and our goals, making it a widely applicable tool. However, the price of this flexibility is that methods designed for a specific system likely outperform our general approach -- future work should explore this possible trade-off. In our work, we extensively investigated how inter-layer structural correlations affect the trade-off between different objectives. We have not, however, explored the scenario when the dynamics of the layers are also directly coupled. Centrality measures that take such coupling into account were only developed for multilayer networks where all layers follow qualitatively similar dynamics~\cite{DeDomenico2015,Sole2016}. It would be interesting to extend rank aggregation techniques to directly consider coupled dynamics.

\begin{acknowledgments} We thank Haochen Wu for the airline network dataset.  We gratefully acknowledge support from the US Army Research Office MURI Award No. W911NF-13-1-0340, DARPA Award No. W911NF-17-1-0077, and the National Institutes of Health Award No. R24-OD011136.
\end{acknowledgments}

\appendix

\section{Local search algorithm}\label{app:LS}

Many methods exist to approximate the Kemeny consensus, a recent survey compared the performance of 104 algorithms and combinations of algorithms~\cite{Ali2012}. It found that so-called local search methods provide an optimal trade-off between accuracy and run-time, meaning that algorithms with significantly longer runtime only marginally decreased the cost function provided by Eq.~(\ref{eq:kemcons}). Local search algorithms start from an initial ranking which can be either random or an approximation of the Kemeny consensus provided by another algorithm. Then this ranking is improved on by a series of local transformations that decrease the cost function. Here, we implement a version of local search based on the simple insert sort algorithm.

As initial ranking we use $R_\T{BC}$, the output of the Borda count algorithm. Let node $i_1$ be the top ranked node in $R_\T{BC}$, node $i_2$ the second, and so on; if there are ties in $R_\T{BC}$, we randomly break them. We then construct a new ranking $R_\T{KA}$ by starting from an empty ranking and iteratively adding nodes. First, we add node $i_1$. We then add $i_2$ such that it has new rank (i)~above $i_1$, (ii)~tied with $i_1$, or (iii)~ranked below $i_1$. The new rank of $i_2$ is chosen to minimize the cost function. We repeat this step until all nodes are assigned a new rank.

\section{Centrality correlated model networks}\label{app:model}

We introduce a method to generate multiplex networks with $N$ nodes, $L$ layers and tunable Kemeny distance $D_{\alpha\beta}$ between each pair of layers $\alpha,\beta\in\{1,2,\dots,L\}$. We start by generating $L$ independent layers using any single-layer network model of choice. We then calculate the rank $r_{\alpha,i}$ of each node $i$ in each layer $\alpha$ based on some centrality $c_{\alpha,i}$. Note that we are not restricted to use the same single-layer network model or the same centrality for all layers. In the following, we describe a procedure to re-label the nodes in each layer to specify the inter-layer dependency between the ranks of nodes.

Let $P(c_1,\dots,c_L)$ be the CDF of the multiplex centrality distribution, i.e., $P(c_1,\dots,c_L)$ is the probability that a randomly selected node $i$ has $c_{\alpha,i}\leq c_\alpha$ for every $\alpha$; and let $P_\alpha(c_\alpha)$ be the CDF of the marginal centrality distribution of layer $\alpha$. Specifying the multiplex centrality distribution would allow us to control the dependency between layers; we, however, cannot arbitrarily choose $P(c_1,\dots,c_L)$, since the marginals $P_\alpha(c_\alpha)$ are determined by the properties of the individual layers. Yet, according to Sklar's theorem, we can always write the multiplex centrality distribution as
\begin{equation}\label{eq:jointCDF}
P(c_1,\dots,c_L) = C(P_1(c_1),\ldots,P_L(c_L)),
\end{equation}
where $C(u_1,\ldots,u_L)$ is an $L$-variate copula~\cite{Sklar1959}. A copula $C:[0,1]^L\rightarrow [0,1]$ is the CDF of a random vector $(u_1,\ldots,u_L)$ with uniform margins~\cite{Joe1997,Nelsen2006}. The advantage of this representation is that it separates the marginal distributions of each layer, specified by $P_\alpha(c_\alpha)$, and the interdependency structure between layers, specified by $C(u_1,\ldots,u_L)$. If the  marginals $P_\alpha(c_\alpha)$ and $P_\beta(c_\beta)$ are continuous, the Kemeny distance $D_{\alpha\beta}$ between the two layers is the same as the Kendall's distance  and it is completely defined by the copula
\begin{equation}\label{eq:D_from_copula}
D_{\alpha\beta} = 1 - 2 \int\limits_0^1 du_\alpha\int\limits_0^1 du_\beta C(u_\alpha,u_\beta) \frac{\partial^2C(u_\alpha,u_\beta)}{\partial u_\alpha\partial u_\beta},
\end{equation}
where $C(u_\alpha,u_\beta)$ is the two-dimensional marginal of the $L$-variate copula~\cite{Nelsen2006}. For example, for independent layers we have $C(u_1,\ldots,u_L)=\prod_{\alpha=1}^Lu_\alpha$, and substituting this into Eq.~(\ref{eq:D_from_copula}), we get $D=0.5$ for all layer pairs.

To re-label the nodes such that the multiplex centrality distribution $P(c_1,\dots,c_L)$ follows Eq.~(\ref{eq:jointCDF}), we draw a random vector $u_i=(u_{1,i},\ldots,u_{L,i})$ from the $L$-variate copula for each node $i$, and from this we obtain a vector of ranks $s_i=(s_{1,i},\ldots,s_{L,i})$ where
\begin{equation}\label{eq:u_rank}
s_{\alpha, i} = N - \lvert \{j\in\{1,\dots,N\}:\,u_{\alpha, i}>u_{\alpha, j}\}\rvert.
\end{equation}
We re-label each node such that $r_{\alpha,i}$, the rank of $i$ in layer $\alpha$, becomes equal to $s_{\alpha,i}$. For example, if $s_i=(1,3)$, we re-label nodes such that the highest ranked node in layer 1 and third highest ranked node in layer 2 are labeled $i$. 

Throughout this paper, we use the Gaussian copula
\begin{equation}\label{eq:gaussian_copula}
C_\rho(u_1,\ldots,u_L) = \Phi_\rho(\Phi^{-1}(u_1),\ldots,\Phi^{-1}(u_L)),
\end{equation}
where $\Phi_\rho$ is the CDF of the L-variate standard normal distribution with correlation matrix $\rho\in [-1,1]^{L\times L}$ and $\Phi^{-1}$ is the inverse CDF of the standard normal distribution. Substituting the Gaussian copula into Eq.~(\ref{eq:D_from_copula}), we find the relationship
\begin{equation}\label{eq:D_to_r}
\rho_{\alpha\beta}= \sin\left(\frac{\pi}{4}(1-D_{\alpha\beta})\right),
\end{equation}
where $\rho_{\alpha\beta}$ is an element of the correlation matrix $\rho$. Therefore, choosing the correlation matrix $\rho$ allows us to set the Kemeny distance between layers.

Finally, note that the above procedure does not take into account ties in the ranks, i.e., we assume that $c_{\alpha,i}\neq c_{\alpha,j}$ for any $i\neq j$. Therefore, if there are any ties in the centralities (e.g., if degree centrality is used), we randomly break them. Furthermore, Eq.~(\ref{eq:D_from_copula}) assumes that the marginal centrality distributions are continuous, if this is not the case, the Kemeny distance $D_{\alpha\beta}$ between layers is no longer exactly provided exactly by Eq.~(\ref{eq:D_from_copula}) and (\ref{eq:D_to_r}); through simulations, however, we found that $D_{\alpha\beta}$ is still well approximated by them.

\subsection{Static scale-free model}

To generate the scale-free layers, we use the static model~\cite{Goh2001}. Starting from $N$ unconnected nodes, we assign a weight $w_i=i^{-\alpha}$ to each node $i=1,\ldots, N$, where $\alpha\in[0,1)$. We then randomly select two nodes $i$ and $j$ with probability proportional to $w_i$ and $w_j$, respectively, and if there is no link between nodes $i$ and $j$, we connect them. We repeat this step until $L$ links are added. The resulting network has average degree $c=2L/N$ and its degree distribution can be written as sum of Poisson distributions
\begin{equation}\label{eq:SFdegdist}
p(k)=\frac{1}{N}\sum\limits_{i=1}^N\frac{c_i^k}{k!} e^{-c_i},
\end{equation}
where $c_i=2Lw_i/(\sum_{j=0}^N w_j)$ is the expected degree of node $i$. For large $N$, the degree distribution is approximated as
\begin{equation}
p(k)=\frac{[c(1-\alpha)]^{1/\alpha}}{\alpha}\frac{\Gamma(k-1/\alpha,c[1-\alpha])}{\Gamma(k+1)},
\end{equation}
where $\Gamma(z)$ is the gamma function and $\Gamma(z,a)$ is the upper incomplete gamma function. We refer to this network as scale-free, because the tail of the distribution decays as a power-law, i.e., $p(k)\sim k^{-\gamma}$, where $\gamma = 1+ 1/\alpha$. 

When numerically evaluating the equations developed in Sec.~\ref{sec:analytic}, we use Eq.~(\ref{eq:SFdegdist}) to represent $p(k)$.

\clearpage
\bibliography{multiplexranking}
\end{document}